\documentclass[draftcls,11pt,onecolumn]{IEEEtran}
\usepackage{times}
\usepackage{amsmath}  
\usepackage{amssymb}  
\usepackage{mathrsfs} 
\usepackage{cite}     
\usepackage{comment}  
\usepackage{upref}
\usepackage{amsfonts}
\usepackage{verbatim}
\usepackage[dvipsnames,usenames]{color}
\usepackage{enumerate}
\usepackage{graphicx}
\usepackage{subfigure}
\usepackage{latexsym}
\usepackage{bm}
\usepackage{multirow}
\usepackage{dsfont}
\usepackage{amsthm,xpatch}
\usepackage{mathtools}
\usepackage{bbm}
\usepackage{latexsym}
\usepackage{accents}
\usepackage{psfrag}
\usepackage{color}
\usepackage{times,color}
\usepackage[usenames,dvipsnames,svgnames,table]{xcolor}
\usepackage{pgf}
\usepackage{tikz}
\usetikzlibrary{arrows, automata, positioning, calc}
\usepackage{cancel} 

\usepackage{algorithm,algpseudocode}

\makeatletter
\newcommand{\removelatexerror}{\let\@latex@error\@gobble}
\makeatletter
\newcommand{\proofpart}[2]{%
	\par
	\addvspace{\medskipamount}%
	\noindent\emph{Part #1: #2}\par\nobreak
	\addvspace{\smallskipamount}%
	\@afterheading
}
\makeatother

\renewcommand{\mathsf}[1]{#1}

\theoremstyle{definition}
\newtheorem{example}{Example}

\usepackage{comment}

\newcommand{\xv}{\ensuremath{\underline{x}}}

\makeatletter
\renewcommand{\paragraph}{%
  \@startsection{paragraph}{4}%
  {\z@}{0.8ex \@plus 0.5ex \@minus .3ex}{-1em}%
  {\normalfont\normalsize\bfseries}%
}
\makeatother

\IEEEoverridecommandlockouts

\begin{document}

\newcommand{\SB}[3]{
\sum_{#2 \in #1}\biggl|\overline{X}_{#2}\biggr| #3
\biggl|\bigcap_{#2 \notin #1}\overline{X}_{#2}\biggr|
}

\newcommand{\Mod}[1]{\ (\textup{mod}\ #1)}

\newcommand{\overbar}[1]{\mkern 0mu\overline{\mkern-0mu#1\mkern-8.5mu}\mkern 6mu}

\makeatletter
\newcommand*\nss[3]{%
  \begingroup
  \setbox0\hbox{$\m@th\scriptstyle\cramped{#2}$}%
  \setbox2\hbox{$\m@th\scriptstyle#3$}%
  \dimen@=\fontdimen8\textfont3
  \multiply\dimen@ by 4             
  \advance \dimen@ by \ht0
  \advance \dimen@ by -\fontdimen17\textfont2
  \@tempdima=\fontdimen5\textfont2  
  \multiply\@tempdima by 4
  \divide  \@tempdima by 5          
  \ifdim\dimen@<\@tempdima
    \ht0=0pt                        
    \@tempdima=\fontdimen5\textfont2
    \divide\@tempdima by 4          
    \advance \dimen@ by -\@tempdima 
    \ifdim\dimen@>0pt
      \@tempdima=\dp2
      \advance\@tempdima by \dimen@
      \dp2=\@tempdima
    \fi
  \fi
  #1_{\box0}^{\box2}%
  \endgroup
  }
\makeatother

\makeatletter
\renewenvironment{proof}[1][\proofname]{\par
  \pushQED{\qed}%
  \normalfont \topsep6\p@\@plus6\p@\relax
  \trivlist
  \item[\hskip\labelsep
        \itshape
    #1\@addpunct{:}]\ignorespaces
}{%
  \popQED\endtrivlist\@endpefalse
}
\makeatother

\makeatletter
\newsavebox\myboxA
\newsavebox\myboxB
\newlength\mylenA

\newcommand*\xoverline[2][0.75]{%
    \sbox{\myboxA}{$\m@th#2$}%
    \setbox\myboxB\null
    \ht\myboxB=\ht\myboxA%
    \dp\myboxB=\dp\myboxA%
    \wd\myboxB=#1\wd\myboxA
    \sbox\myboxB{$\m@th\overline{\copy\myboxB}$}
    \setlength\mylenA{\the\wd\myboxA}
    \addtolength\mylenA{-\the\wd\myboxB}%
    \ifdim\wd\myboxB<\wd\myboxA%
       \rlap{\hskip 0.5\mylenA\usebox\myboxB}{\usebox\myboxA}%
    \else
        \hskip -0.5\mylenA\rlap{\usebox\myboxA}{\hskip 0.5\mylenA\usebox\myboxB}%
    \fi}
\makeatother

\xpatchcmd{\proof}{\hskip\labelsep}{\hskip3.75\labelsep}{}{}

\pagestyle{plain}

\title{\fontsize{21}{28}\selectfont On Accelerated Testing for COVID-19\\ Using Group Testing}

\author{Krishna R. Narayanan, Anoosheh Heidarzadeh, and Ramanan Laxminarayan\thanks{K. R. Narayanan and A. Heidarzadeh are with the Department of Electrical and Computer Engineering, Texas A\&M University, College Station, TX 77843 USA (E-mail: \{krn,anoosheh\}@tamu.edu).}\thanks{R. Laxminarayan is with the Center for Disease Dynamics, Economics \& Policy, Silver Spring, MD 20910, USA, the Department of Global Health, University of Washington, Seattle, WA 98104, USA, and the Princeton Environment Institute, Princeton University, Princeton, NJ 08544, USA (E-mail: ramanan@cddep.org).}}



\maketitle 

\thispagestyle{plain}

\begin{abstract}
COVID-19 has resulted in a global health crisis that may become even more acute over the upcoming months. One of the main reasons behind the current rapid growth of COVID-19 in the U.S. population is the limited availability of testing kits and the relatively-high cost of screening tests. In this draft, we demonstrate the effectiveness of group testing (pooling) ideas to accelerate testing for COVID-19. This draft is semi-tutorial in nature and is written for a broad audience with interest in mathematical formulations relevant to COVID-19 testing. Therefore, ideas are presented through illustrative examples rather than through purely theoretical formulations. The focus is also on pools of size less than 64 such as what is practical with current RT-PCR technology.
\end{abstract}

\section{Introduction}
Epidemiologists believe large-scale testing and isolation of infected people 
is among the most effective strategies to control the spread of COVID-19. 
One of the major reasons testing has been substantially delayed in many countries including the United States is the limited availability of screening tests \cite{khazan_2020}. 
The resulting inability to rapidly test large sections of the U.S. population for COVID-19 is exacerbating the health crisis. 
While more testing kits have started to become available within the last few days, it is unlikely that the number of available tests will
scale fast enough to be able to test large sections of the population rapidly.
Here, we investigate the use of ideas from group testing (pooling) to accelerate testing for COVID-19. 

The COVID Tracking Project provides real-time data on the number of tests conducted in each state in the United States and the number of positive tests among them. This data for a few states and for the whole country is shown in Table~\ref{table:data1}. 
\begin{table}[h]
    \centering
    \caption{Number of tests and positives in each state as of March 22, 2020.}
    \begin{tabular}{|c|c|c|c|}
    \hline
       State        & Number of tests   & Number of positives   & \% Positives \\
    \hline
        Washington  & 27,121            & 1,793                  & 6.61\%          \\
    \hline
        California  & 12,840            & 1,536                  & 11.96\%         \\
    \hline
        Texas       & 8,756              & 334                   & 3.81\%         \\
    \hline
       U.S.A        & 225,374           & 31,888                & 14.14\%           \\
     \hline
    \end{tabular}
    \label{table:data1}
\end{table}{}
The are two main takeaways from this table: (i) only a \emph{small number} of people are being tested in each state, and (ii) only a {\em small fraction} of the people who were tested for COVID-19 tested positive. In this regime, the group testing is expected to yield a significant reduction in the number of tests per person, as compared to performing one test on each individual separately.

One of the effective strategies for managing the spread of COVID-19 is large-scale population-level testing of asymptomatic people for obtaining coarse grained information. In such cases, the probability of a person who is tested being infected (prevalence) will be even smaller, increasing the efficiency of group testing. 

In the current context, group testing can be useful for two problems:
\begin{enumerate}
    \item \emph{Testing individuals:} Group testing can be used to decrease the number of tests required to identify infected individuals within a population.
    \item \emph{Testing populations:} Group testing can be used to classify the infection rate in a neighborhood as being high or low, i.e., group testing can be used in conjunction with hypothesis testing.
\end{enumerate}

\section{Group Testing}
Consider the problem of testing a population of $n$ people where each person in the population is likely to have been infected with probability $p$, so as to determine the subset of infected people. 
The traditional way to determine the set of infected people is to conduct individual tests on swabs obtained from each person in the population. This requires $n$ tests or, one test per person.

Group testing is an alternative way to solve the problem, which in some cases, requires drastically fewer than $n$ tests. 
The key idea in group testing is to group (or, pool) multiple items (swabs or blood samples) and test each group rather than each individual.
The output of the test will be \textit{negative} if everyone in the group is healthy or else, the output will be \textit{positive.} 
The objective of group testing is to design the testing scheme (or, pools) such that the total number of tests $m$ to be performed is minimized.
The remarkable result from group testing is that if $p$ is small, the infected people can be identified from far fewer than $n$ tests, i.e., the average number of tests per person can be substantially smaller than one. Group testing was first introduced to the field of statistics by Dorfman \cite{dorfman1943detection} during World War II for testing soldiers for syphilis without having to test each soldier individually. 
Since then, group testing has been used in 
such as clone library screening, non-linear optimization, multi-access communication etc.., \cite{du1999combinatorial} 
and fields like biology\cite{chen2008survey}, machine learning\cite{malioutov2013exact}, data structures\cite{goodrich2005indexing} and signal processing~\cite{emad2014poisson}.
In turn, advances in coding theory and multi-access communication have substantially advanced the field of group testing. A comprehensive survey of group testing algorithms can be found in \cite{AJS2019,du1999combinatorial,chan2014non,atia2012boolean,Mazumdar2016,Barg2017}. 


There are three versions of group testing that are commonly used. In non-adaptive group testing, the pools are formed before testing begins and typically, tests on the pools are conducted in parallel. In adaptive group testing, pools are formed adaptively, one at a time, after observing the results of tests on earlier pools. Adaptive group testing, obviously, is more effective in terms of number of tests required; however, it can be slow since each test may take hours to complete. Multi-stage group testing is a compromise between the fully adaptive and fully non-adaptive versions. Here there are $L$ rounds of tests and the pools during the $i$th round can be formed after observing the results of tests from pools formed until stage $i-1$.

\section{Determining Infected Individuals Using Group Testing}
In this section, we will explain a few group testing schemes and demonstrate their effectiveness through some examples. 

\begin{figure}[t]
    \centering
    \includegraphics[width=3.5in]{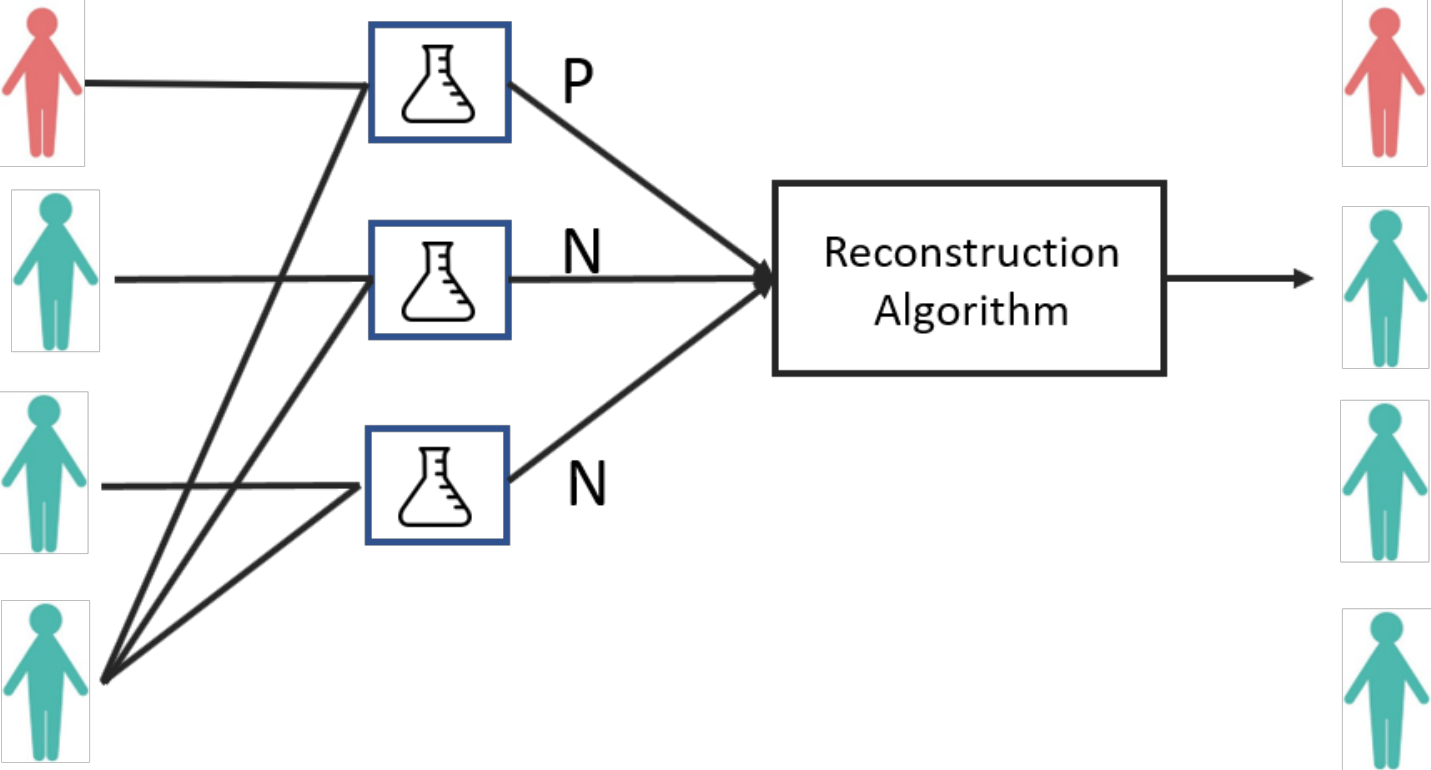}
    \caption{A schematic showing a non-adaptive group testing strategy for $n=4$ people.}
    \label{fig:GTschematic}
\end{figure}

\begin{example}{\bf Non-adaptive group testing:} 
\label{example:nonadaptive4}Consider a population of $n=4$ people (labeled $1,2,3,4$), among whom at most $k=1$ people are infected. An obvious way to determine the infected people would be to run a test on each individual, requiring $4$ tests in total. However using the idea of group testing, one can pool $m=3$ different subsets of $n=4$ people, and run \emph{one test} on each pool, and determine the $k=1$ infected people (if any). For instance, consider the $m=3$ pools $P_1=\{1,4\}$, $P_2=\{2,4\}$, and $P_3=\{3,4\}$. The selected pools in this example can be represented by a $3\times 4$ binary matrix, called a \emph{testing matrix}, as follows: \[M = \begin{bmatrix}1 & 0 & 0 & 1\\ 0 & 1 & 0 & 1\\ 0 & 0 & 1 & 1\end{bmatrix},\] 
where each row corresponds to one test, and the entries $1$ in that row represent the people corresponding to that test. For example, the first row of matrix $M$ represents a test on the pool of individuals $1$ and $4$. 
The main property of the matrix $M$ is that \emph{every column is distinct} and \emph{not all-zero}. Such a matrix is called \emph{$\bar{1}$-separable} in the group testing literature. (In general, a binary matrix is called \emph{$k$-separable} (or \emph{$\bar{k}$-separable}) if no $k$  columns (or no $k$-or-fewer columns) have the same Boolean sum (bitwise OR).) Using the $\bar{1}$-separability of the testing matrix $M$, the results of the $m=3$ tests can be used to determine the $k=1$ infected person (if any). For instance, suppose that the vector of results of the $m=3$ tests is $[1,0,0]^{\mathsf{T}}$, i.e., the first test is positive (represented by Boolean $1$), the second test is negative (represented by Boolean $0$), and the third test is negative ($0$). Then, we will determine that the first person is infected. As an another instance, if the vector of the test results is $[0,0,0]^{\mathsf{T}}$, we will determine that none of the people is infected.     
A schematic showing this testing procedure is shown in Fig.~\ref{fig:GTschematic}.
\end{example}

\begin{example}
\label{example:adaptive4}
{\bf~Adaptive group testing:} Consider the same scenario as in Example~\ref{example:nonadaptive4}, i.e., a population of $n=4$ people (labeled $1,2,3,4$), among whom at most $k=1$ people are infected. In adaptive group testing, we can first form the pool $P_1 = \{1,2\}$ and test the pool. If the result is positive, then we use the pool $P_2 = \{1\}$. If the test on pool $P_1$ is negative, then we test the pool $P_3 = \{3,4\}$ and if the result is positive, then we test $P_4 = \{3\}$. It can be seen that this procedure identifies the infected person (if there is one) with fewer tests on the average than that in the non-adaptive approach. However, in the worst case, we still require 3 tests and more importantly, since the tests are performed sequentially, the total time taken can be three times as long as in the non-adaptive case.
\end{example}

\begin{example}\label{example:SOMS}
{\bf~Sparsity-oblivious multi-stage (SOMS) group testing:} Consider a population of $n=4$ people (labeled $1,2,3,4$). An obvious way to determine the infected people would be to run a test on each individual, requiring $4$ tests in total. However, with group testing, we can follow the procedure shown in the flowchart in Figure~\ref{fig:flowchart}. We refer to this strategy as the \emph{Sparsity-Oblivious Multi-Stage (SOMS) strategy}. 
Here $T_i$ refers to the $i$th test and $\{a,b,c\}$ refers to pooling the samples of $a,b$ and $c$. For example, $T_1:\{1,2,3,4\}$ means that the first test is conducted on a pool consisting of all 4 people. If the result of that test is negative (denoted by $\mathrm{N}$), we stop since it means no one is infected. If the result of that test is positive (denoted by $\mathrm{P}$), then we conduct three more tests $T_2,T_3$ and $T_4$ on pools $\{1,2\}$, $\{3,4\}$, and $\{1,3\}$, respectively. 
\begin{figure}
    \centering\vspace{-0.45cm}
    \includegraphics[width=3in]{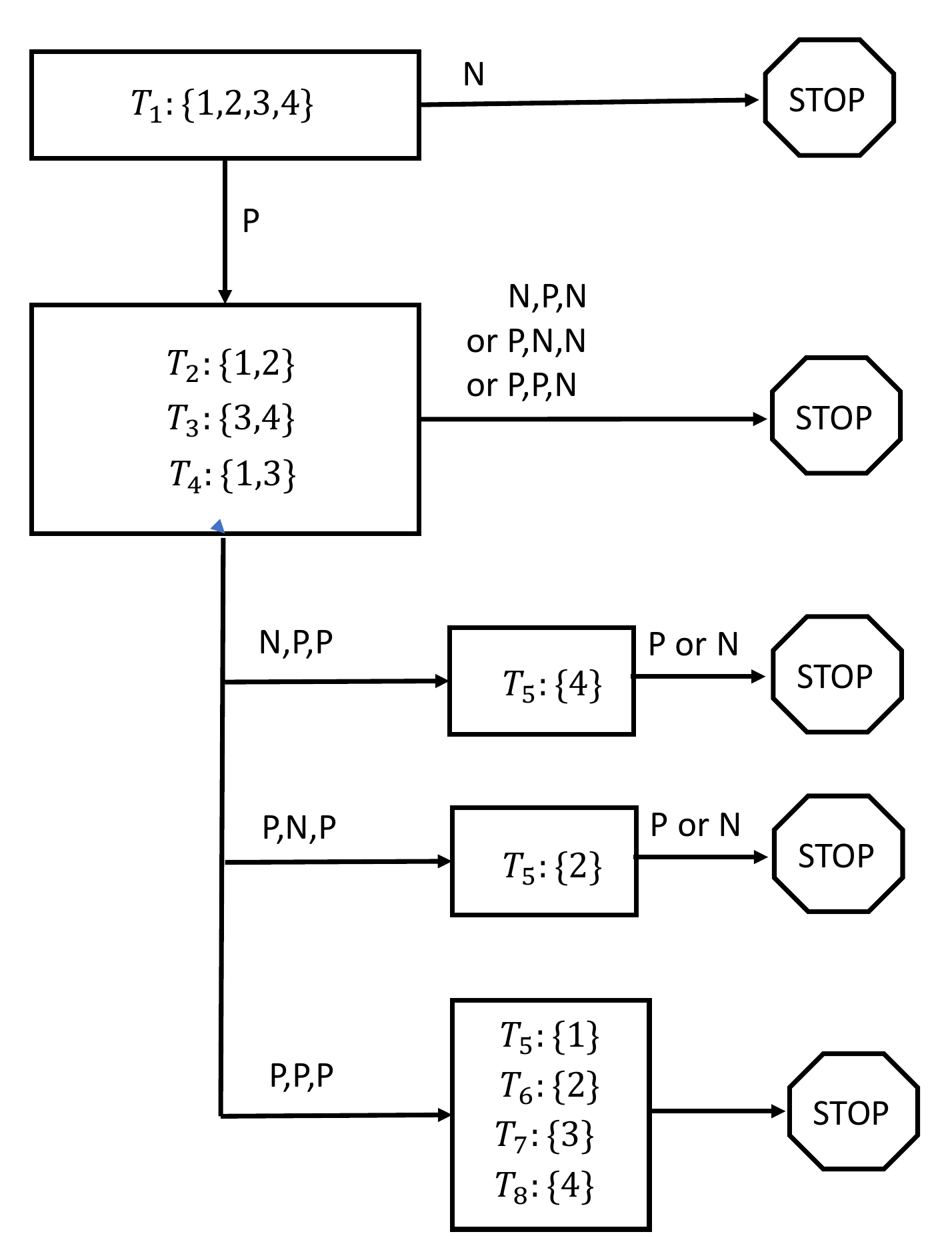}\vspace{-0.125cm}
    \caption{Flowchart showing the SOMS group testing strategy for $n=4$ people.}
    \label{fig:flowchart}
\end{figure}
If the result of the three tests $T_2,T_3$ and $T_4$ are $(\mathrm{N},\mathrm{P},\mathrm{N})$, $(\mathrm{P},\mathrm{N},\mathrm{N})$ or $(\mathrm{P},\mathrm{P},\mathrm{N})$, then we can uniquely identify the one person who is infected and we can stop. 
If the results of $T_2,T_3$ and $T_4$ are $(\mathrm{N},\mathrm{P},\mathrm{P})$, respectively, we conduct a fifth test just on $\{4\}$ and we stop. 
If the results of $T_2,T_3$ and $T_4$ are $(\mathrm{P},\mathrm{N},\mathrm{P})$, respectively, then we conduct a fifth test just on $\{2\}$ and we stop. 
If the results of $T_2,T_3$ and $T_4$ are $(\mathrm{P},\mathrm{P},\mathrm{P})$, respectively, then we conduct four tests individually on all 4 samples.
\end{example}

\begin{example}\label{example:SOFA}
{\bf~Sparsity-oblivious fully-adaptive (SOFA) group testing:} One of the main advantages of the SOMS strategy is that it leverages parallel tests at some stages of the testing. This can significantly reduce the \emph{total testing execution time} in order to identify all infected people in the population (on average). However, this advantage comes at the price of not being optimal in terms of the \emph{total number of tests}. In what follows, we propose another testing strategy, referred to as the \emph{Sparsity-Oblivious Fully-Adaptive (SOFA) strategy}, that will require less number of tests on average for a population of $n=4$ people. It should be noted that the average execution time of SOFA is longer than that of SOMS. That is, the SOMS and SOFA strategies achieve different tradeoffs between the average number of tests and the average execution time.  

In SOFA, depending on the results of the previous tests, at each stage either (i) we perform one test on the pool of all unidentified people, or (ii) we perform a binary search on them until we find one (and no more) infected person (if any). 
In order to choose the best action ((i) or (ii)) at each stage, the algorithm uses the number of infected people that have already been identified.
When the number of already-identified infected people is relatively larger (or smaller) than the expected number of infected people ($np$), the action (i) (or (ii)) will be taken. This procedure is motivated by the fact that when the number of already-identified infected people is relatively large as compared to the expected number of infected people, most likely none of the remaining people are infected; and hence it would be more efficient (in terms of the number of tests) to perform one test on all unidentified people, instead of performing a binary search on them.    

The main advantage of any efficient group testing strategy is that the average number of tests required can be substantially lower than $n$. Let $p$ be the probability that each person who appears for the test is infected, independently from other people. 

\begin{figure}
    \centering\vspace{-0.25cm}
    \includegraphics[width=3.5in]{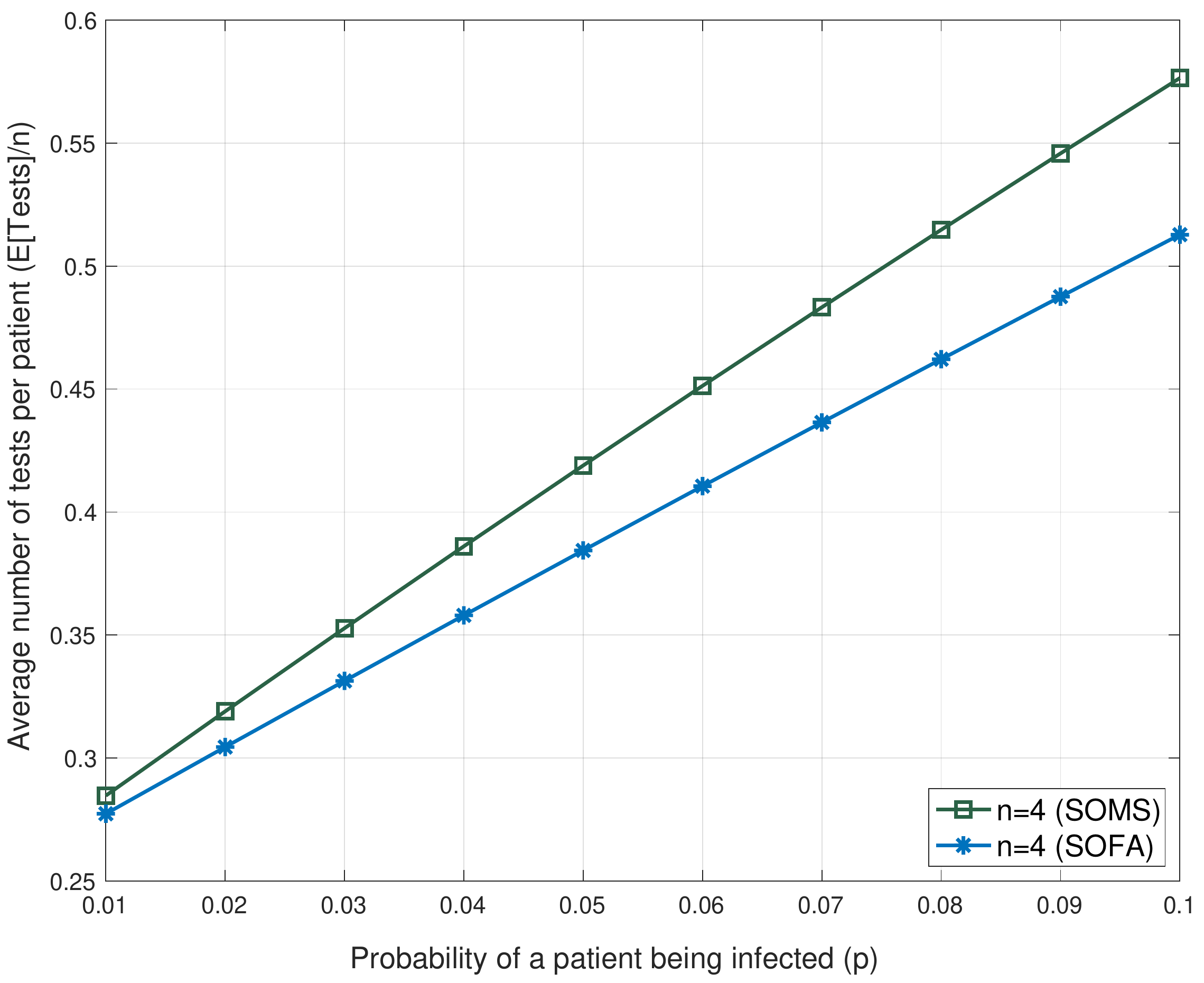}\vspace{-0.25cm}
    \caption{Average number of tests per person for $n=4$ and different values of $p$.}
    \label{fig:results1}
\end{figure}
In Figure~\ref{fig:results1}, a plot of the average number of required tests per person is shown for $n=4$ people and for different values of $p$, when using both the SOMS and SOFA strategies. It can be seen that when $p=0.05$, the average number of tests required for identifying all infected people in a population of $n=4$ people using SOMS (or SOFA) is about $0.425\times n = 1.7$ (or $0.375\times n = 1.5$). 
This shows a reduction in the number of tests by about $57.5\%$ for SOMS (or $62.5\%$ for SOFA), as compared to testing the people individually. 
Not surprisingly, when $n$ is fixed, the average number of required tests increases as $p$ increases. That said, when $n=4$, even for $p=0.1$, a reduction of about $42.5\%$ (or $50\%$) can be achieved using SOMS (or SOFA). 

\begin{figure}
    \centering
    \includegraphics[width=3.5in]{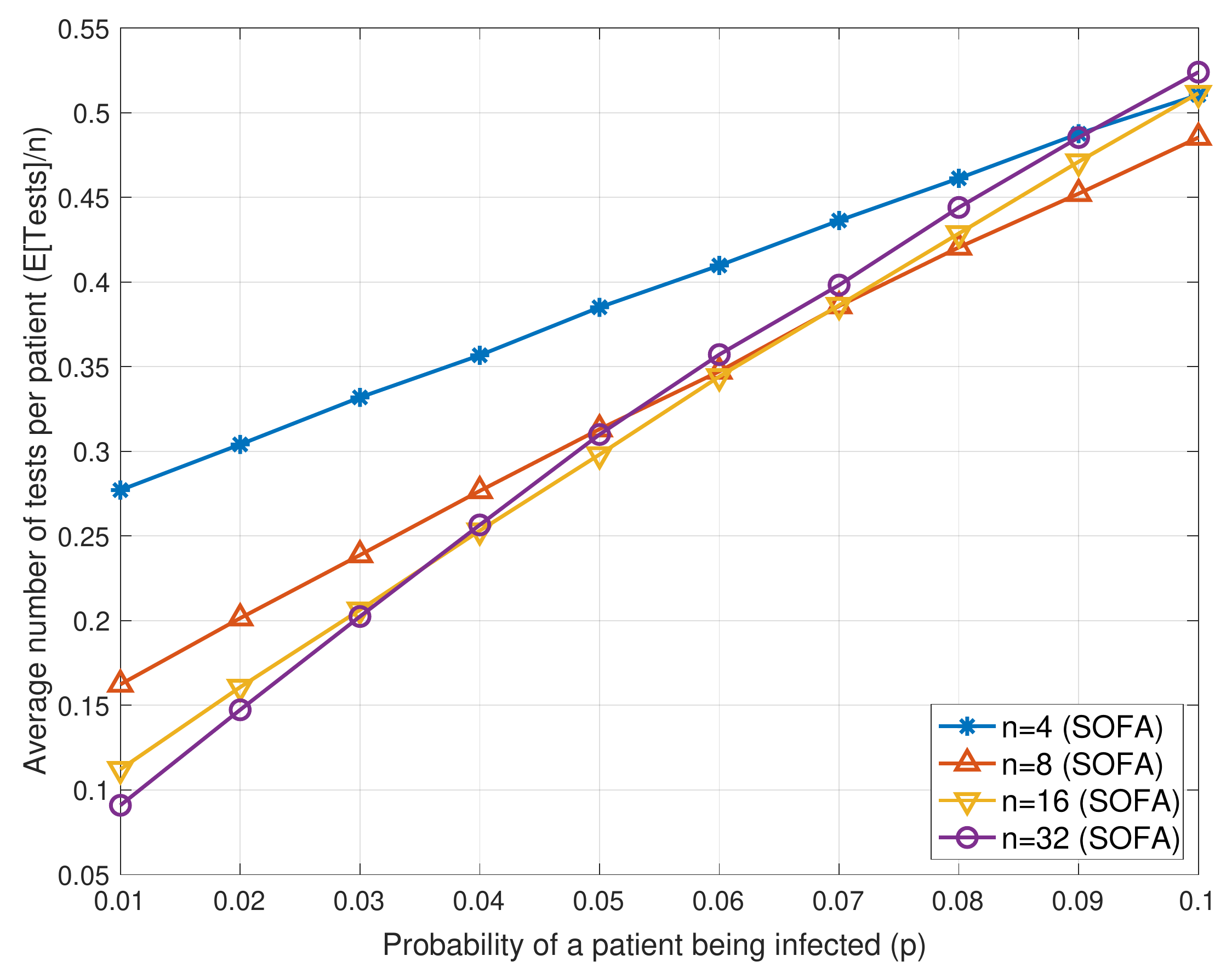}\vspace{-0.25cm}
    \caption{Average number of tests per person for different values of $n$ and $p$.}
    \label{fig:results2}
\end{figure}
Figure~\ref{fig:results2} depicts a plot of the average number of tests per person when using SOFA for $n\in \{4,8,16,32\}$ people and different values of $p$. As can be seen, for sufficiently small values of $p$, the reduction in the average number of tests (when compared to individual testing) becomes even more profound for larger values of $n$. 
For instance, using SOFA, when $p=0.01$, the average number of tests for $n=32$ people is about $0.1\times n = 3.2$, resulting in about $90\%$ reduction in the number of tests; whereas for $n=4$ people the reduction is about $72.5\%$. 
\end{example}

\begin{figure}
    \centering
    \includegraphics[width=3.5in]{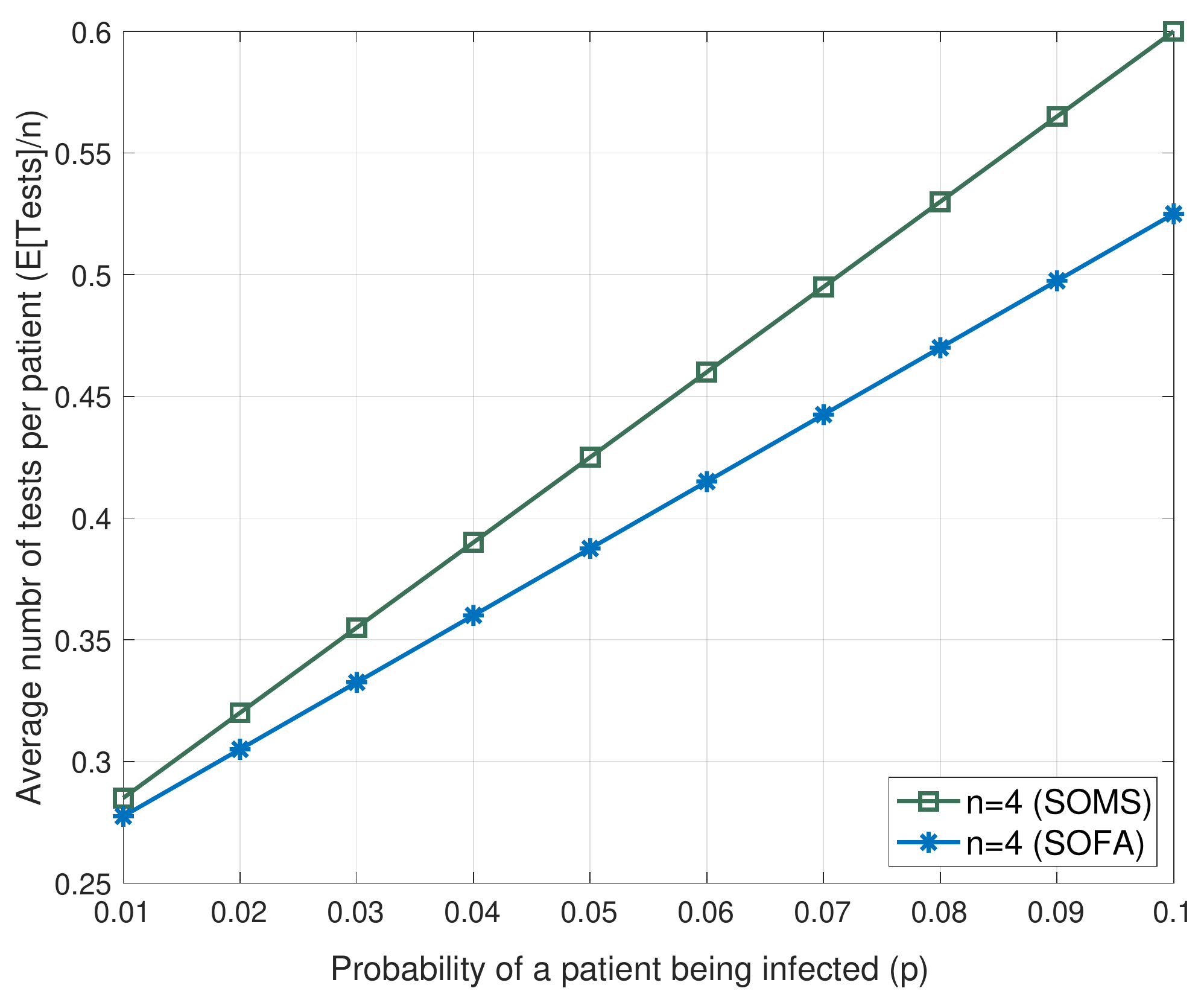}\vspace{-0.25cm}
    \caption{Average number of tests per person for the worst-case correlation, for $n=4$ and different values of $p$.}
    \label{fig:results3}\vspace{-0.25cm}
\end{figure}

\subsection{Impact of correlation}
A natural question to ask is how the correlation between the infection rates of people in the population affects the performance of the group testing algorithms. Equivalently, what would the performance of the group testing algorithms under worst-case correlation structure? Let $X_k \sim \mbox{Bernoulli}(p), k=1,\ldots,n$ denote binary random variables representing the infection status of each person in the sample. 
The infection status of the entire group is then given by a vector-valued random variable $\underline{X} = [X_1, X_2, \ldots, X_n] \in \{0,1\}^n$. Let $\underline{x}$ be a realization of $\underline{X}$.
Let $Z(\underline{x})$ denote the integer representation of $\underline{x}$ and let
$[\underline{x}]_k$ denote the $k$th bit of $\underline{x}$. 
Let $\mathcal{B}_k := \{i : [Z^{-1}(i)]_k=1\}$ be the set of integers $i$ such that the $k$th bit in the binary representation of $i$ is 1.
Let $\Gamma_i$ denote the number of tests required to determine $Z^{-1}(i)$ correctly.
Any arbitrary correlation between $X_k$'s can be captured by defining a distribution on $\underline{X}$.
Let $\underline{\pi} = [\pi_1, \ldots, \pi_{2^n}]$ denote the probability mass function of $\underline{X}$, i.e.,
$\pi_i = \mathbb{P}(Z(\underline{x})=i)$. We can solve for the worst case expected value of $\Gamma$ over $\underline{\pi}$
by solving the following linear program
\begin{eqnarray}
\nonumber
\text{minimize:} - \sum_{i=1}^{2^n} \Gamma_i \pi_i & & \\
\nonumber
\text{Subject to:}~-\pi_i & \leq & 0 \\
\nonumber
\sum_i \pi_i & = & 1 \\
\label{eqn:LPmarginalconstraints}
\sum_{j \in \mathcal{B}_k} \pi_j & = & p, \ \ \forall k = 1, \ldots, n
\end{eqnarray}

The constraints in \eqref{eqn:LPmarginalconstraints} refer to the constraints on the marginal probabilities of each $X_k$ being $p$.
The performance of both the SOMS and SOFA strategies for the worst-case correlation subject to the constraint that the probability of infection of any individual is $p$, is plotted in Figure~\ref{fig:results3} for $n=4$ and different values of $p$. As can be seen, even for the worst-case correlation, the average number of tests for both the SOMS and SOFA strategies are substantially lower than that for individual testing.

\section{Group Testing for Infection Rate Classification}

We now move from the problem of using group testing for testing individuals to the problem of using group testing for obtaining coarse-grained information about the prevalence of infections in neighborhoods. 
Consider the problem of classifying the infection rate in a neighborhood or geographic region as being low $(p_0)$ or high $(p_1)$ when $p_0$ and $p_1$ are known {\em a priori}. Group testing can be very beneficial in reducing the number of tests required for this classification task. This is a hypothesis testing problem with two hypotheses
\[
\begin{array}{cl}
H_0 :    & \hbox{Infection rate is}~p_0 \\
H_1 :    & \hbox{Infection rate is}~p_1
\end{array}
\]

In our approach, we randomly select a group of $N$ people from the region and split them into $L$ subpools, $S_1,S_2,\ldots,S_L$, each of size $N/L$. A subpool $S_k$ is said to be infected if at least one person in the subpool is infected. Let $X_k$ be a binary random variable defined as follows
\[
X_k = \left\{
  \begin{array}{ll}
    1, & \mbox{if $S_k$ is infected;} \\
    0, & \mbox{if $S_k$ is not infected.}
  \end{array}
\right.
\]
We assume a uniformly random sampling process such that each person in the pool is infected with probability $p_i$ independent of every other person in the chosen samples. Therefore, $X_i$ is a Bernoulli random variable with parameter $q_i$ if hypothesis $H_i$ is true, where $q_i$ is given by
\[
q_i = 1-(1-p_i)^{N/L}.
\]

Let $\xv = [x_1,x_2,\ldots,x_L]$ denote a vector of realizations of $X_k$s indicating which subpools have been infected and let $n(\xv):=\sum_k x_k$ denote the number of infected subpools. If the prior probabilities of $H_0$ and $H_1$ are given by $\pi_0$ and $\pi_1$, respectively, the log likelihood ratio (LLR) for $\xv$ is given by
\begin{eqnarray}
\label{eqn:llr} \nonumber 
    L(\xv) & := & \log \frac{\mathbb{P}(H_0|\xv)}{\mathbb{P}(H_1|\xv)} \\
    \nonumber & = & \log \frac{\pi_0}{\pi_1} + n(\xv) \log \frac{q_0}{q_1} + (L-n(\xv)) \log \frac{1-q_0}{1-q_1} \\
    & = & \log \frac{\pi_0}{\pi_1} + n(\xv) \bigg(\log \frac{q_0}{q_1} - \log \frac{1-q_0}{1-q_1} \bigg) + L \log \frac{1-q_0}{1-q_1}. 
\end{eqnarray}
It is clear from \eqref{eqn:llr} that $n(\xv)$ is a sufficient statistic for the LLR test. Hence, the solution to the hypothesis testing problem is given by 
\begin{equation}
\label{eqn:computeV}
    \mbox{Select}~H_0~\mbox{if}~ n(\xv) \leq V = 
    \Biggl\lfloor \frac{\log\frac{\pi_0}{\pi_1}+L \log\frac{1-q_0}{1-q_1}}{-\log\frac{q_0}{q_1} + \log\frac{1-q_0}{1-q_1}} \Biggr\rfloor.
\end{equation}

The main idea in our approach is to perform group testing using binary splitting at the subpool level to determine whether $n(\xv) \leq V$. 
An important difference between our algorithm and conventional binary splitting is that 
we do not perform binary splitting until we recover $\xv$ exactly; 
rather, we perform binary splitting only until we are able to ascertain if $n(\xv) \leq V$. 
Since $\mathbbm{1}_{n(\xv) \leq V}$ is a function of $\xv$, 
our algorithm will typically require fewer tests on the average compared to conventional binary splitting.

\subsection{Probability of false alarm, probability of detection and average number of tests}
The probability of false alarm ($P_F$) and probability of correct detection ($P_D$) are given by
\begin{eqnarray}
P_F & := & \mathbb{P}(n(\xv) > V | H_0~\mbox{is true}) \nonumber \\ 
&=& \sum_{j=V+1}^L \binom{L}{j} q_0^j (1-q_0)^{L-j}, \nonumber \\[1em]
P_D & := & \mathbb{P}(n(\xv) > V | H_1~\mbox{is true}) \nonumber \\ 
&=& \sum_{j=V+1}^L \binom{L}{j} q_1^j (1-q_1)^{L-j}.\nonumber
\end{eqnarray}

For every $\xv$, let $\Gamma(\xv)$ denote the number of tests required. 
Since $\xv$ is a random vector, $\Gamma(\xv)$ is a random variable and the average number of tests is given by
$\mathbb{E}[\Gamma] = \sum_{\xv} \mathbb{P}(\xv) \Gamma(\xv)$.
The following example will explain the group testing procedure and an analysis of the procedure in more detail. 

\begin{figure}
    \centering
    \includegraphics[width=4in]{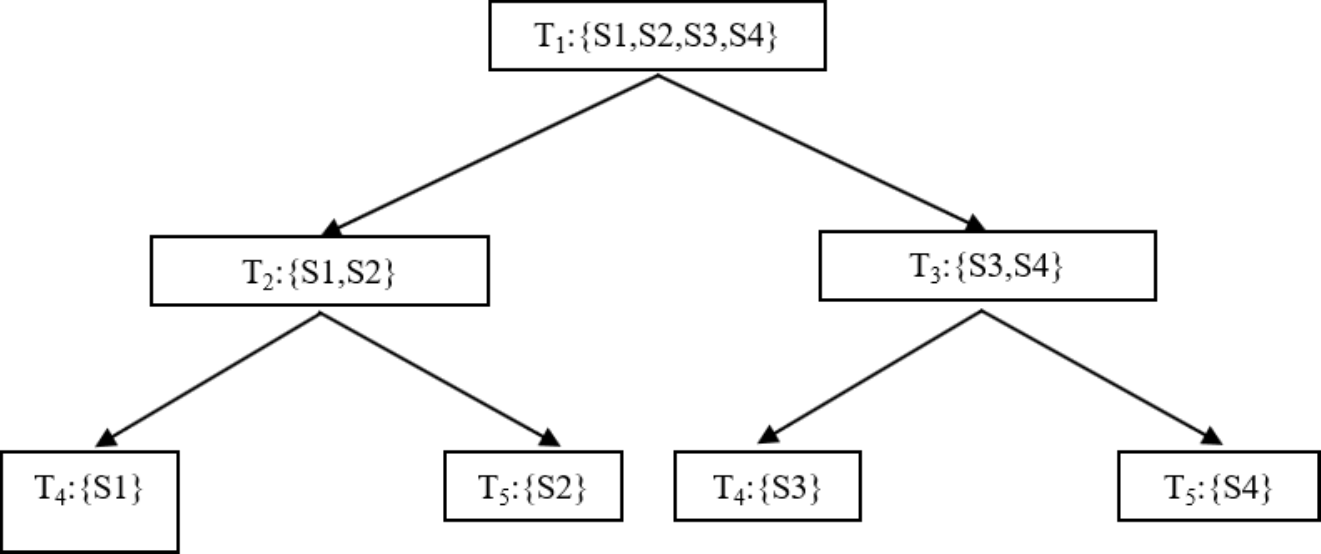}
    \caption{Flowchart representing hypothesis testing using binary splitting.}
    \label{fig:flowchartbinarysplitting}
\end{figure}

\begin{example}
Suppose $p_0 = 0.01,p_1 = 0.05,\pi_0=0.5,\pi_1=0.5$. We choose $N=64$ people from a region and form $L=4$ subpools labeled $S_1,S_2,S_3,S_4$. $V$ can be computed from \eqref{eqn:computeV} and for these parameters, $V=1$. Our group testing procedure is shown in Figure~\ref{fig:flowchartbinarysplitting}.
\end{example}
We use the notation $T_i = \{S_a,S_b,S_c\}$ to denote that the $i$th test is performed by pooling all the samples from the subpools $a,b,$ and $c$.
$T_1:\{S_1,S_2,S_3,S_4\}$ then refer to performing one test by pooling the samples from subpools $S_1,S_2,S_3$, and $S_4$. If the result of test $T_1$ is negative, we accept $H_0$. If the result of $T_1$ is positive, we perform two more tests $T_2:\{S_1,S_2\}$ and $T_3:\{S_3,S_4\}$. If the results of both tests $T_2$ and $T_3$ are positive, we accept $H_1$. 
If the result of only one of $T_1$ or $T_2$ is positive, then we perform two more tests by breaking that pool into two smaller subpools. For example, if the result of $T_2$ is positive, then we perform two more tests $T_4:\{S_3\}$ and $T_5:\{S_4\}$.  If the result of only one of $T_4$ or $T_5$ is positive, then we accept $H_0$, else we accept $H_1$. 
We can analyze the probability of false alarm, probability of detection, and the average number of tests required as follows. The distribution of the number of infected subpools, the chosen hypothesis, and the number of tests required in each case are shown in Table~\ref{table:distributionL=4} for $L=4$.

\begin{table}[]
    \centering
    \caption{Chosen hypothesis and number of tests needed for different $\xv$ vectors.}
    \begin{tabular}{|c|c|c|c|}
    \hline
    $\xv$ & $\mathbb{P}(\xv|H_i)$ & Chosen hypothesis & No. of tests $\Gamma(\xv)$\\
    \hline
    0 0 0 0  & $(1-q_i)^4$   & $H_0$ & 1 \\
    \hline
    0 0 0 1   & $q_i(1-q_i)^3$   & $H_0$ & 5 \\
    \hline
    0 0 1 0   & $q_i(1-q_i)^3$   & $H_0$ & 5 \\
    \hline
    0 0 1 1   & $q_i^2(1-q_i)^2$   & $H_1$ & 5 \\
    \hline
    0 1 0 0   & $q_i(1-q_i)^3$   & $H_0$ & 5 \\
    \hline 
    0 1 0 1   & $q_i^2(1-q_i)^2$   & $H_1$ & 3 \\
    \hline 
    0 1 1 0   & $q_i^2(1-q_i)^2$   & $H_1$ & 3 \\
    \hline 
    0 1 1 1   & $q_i^3(1-q_i)$   & $H_1$ & 3 \\
    \hline 
    1 0 0 0  & $q_i(1-q_i)^3$   & $H_0$ & 5 \\
    \hline
    1 0 0 1   & $q_i^2(1-q_i)^2$   & $H_1$ & 3 \\
    \hline
    1 0 1 0   & $q_i^2(1-q_i)^2$   & $H_1$ & 3 \\
    \hline
    1 0 1 1   & $q_i^3(1-q_i)$   & $H_1$ & 3 \\
    \hline
    1 1 0 0   & $q_i^2(1-q_i)^2$   & $H_1$ & 5 \\
    \hline 
    1 1 0 1   & $q_i^3(1-q_i)$   & $H_1$ & 3 \\
    \hline 
    1 1 1 0   & $q_i^3(1-q_i)$   & $H_1$ & 3 \\
    \hline 
    1 1 1 1   & $q_i^4$   & $H_1$ & 3 \\
    \hline 
    \end{tabular}
    \label{table:distributionL=4}
\end{table}

The probability of false alarm ($P_F$) is the probability that 2,3, or 4 subpools are infected when 
$H_0$ is true and the probability of detection ($P_D$) is the probability that 2,3, or 4 subpools are infected when $H_1$ is true. These probabilities are given by
\begin{eqnarray}
P_F & = & 6q_0^2 (1-q_0)^2+4q_0^3 (1-q_0 )+q_0^4, \nonumber \\
P_D & = & 6q_1^2 (1-q_1)^2+4q_1^3 (1-q_1 )+q_1^4. \nonumber
\end{eqnarray}
The average number of tests required can be computed by taking the expected value of the random variable $\Gamma$ given by
\begin{equation*}
    \mathbb{E}[\Gamma]  =  \big[(1-q_i)^4\big] \cdot 1 + \big[4 q_i^2 (1-q_i)^2 + 4 q_i^3 (1-q_i) + q_i^4\big] \cdot 3 
    + \big[4 q_i (1-q_i)^3 + 2 q_i^2 (1-q_i)^2 \big] \cdot 5.
\end{equation*}{} 

\vspace{-1cm}
\subsection{Extensions}
The above algorithm can be generalized in many ways. A proper choice of $L$ is important to obtain several points in the $(P_F,P_D)$-plane also called as the receiver operating characteristic (ROC) curve. 
As is common in any binary hypothesis testing, we can change the threshold $V$ to trade off $P_F$ for $P_D$.
When $N$ is large, the result of $T_1$ will be positive with high probability when either $H_0$ or $H_1$ is true. In this case, $T_1$ will not be very informative. Hence, we can skip this test and directly start with tests at the next level, namely $T_2$ and $T_3$. This strategy will reduce the number of tests without affecting the probability of false alarm and the probability of detection significantly. 
The average number of tests in this case is given by
\begin{equation*}
    \mathbb{E}[\Gamma]  = \big[(1-q_i)^4 + 4 q_i^2 (1-q_i)^2 + 4 q_i^3 (1-q_i) + q_i^4\big] \cdot 2 + \big[4 q_i (1-q_i)^3 + 2 q_i^2 (1-q_i)^2 \big] \cdot 4. 
\end{equation*}
More generally, we can directly start the tests at level $\tau$, and $\tau$ can be tuned as a parameter.


\subsection{Results}
\subsubsection{Noiseless Setting}
\begin{figure}[t]
    \centering
    \includegraphics[width=3.75in]{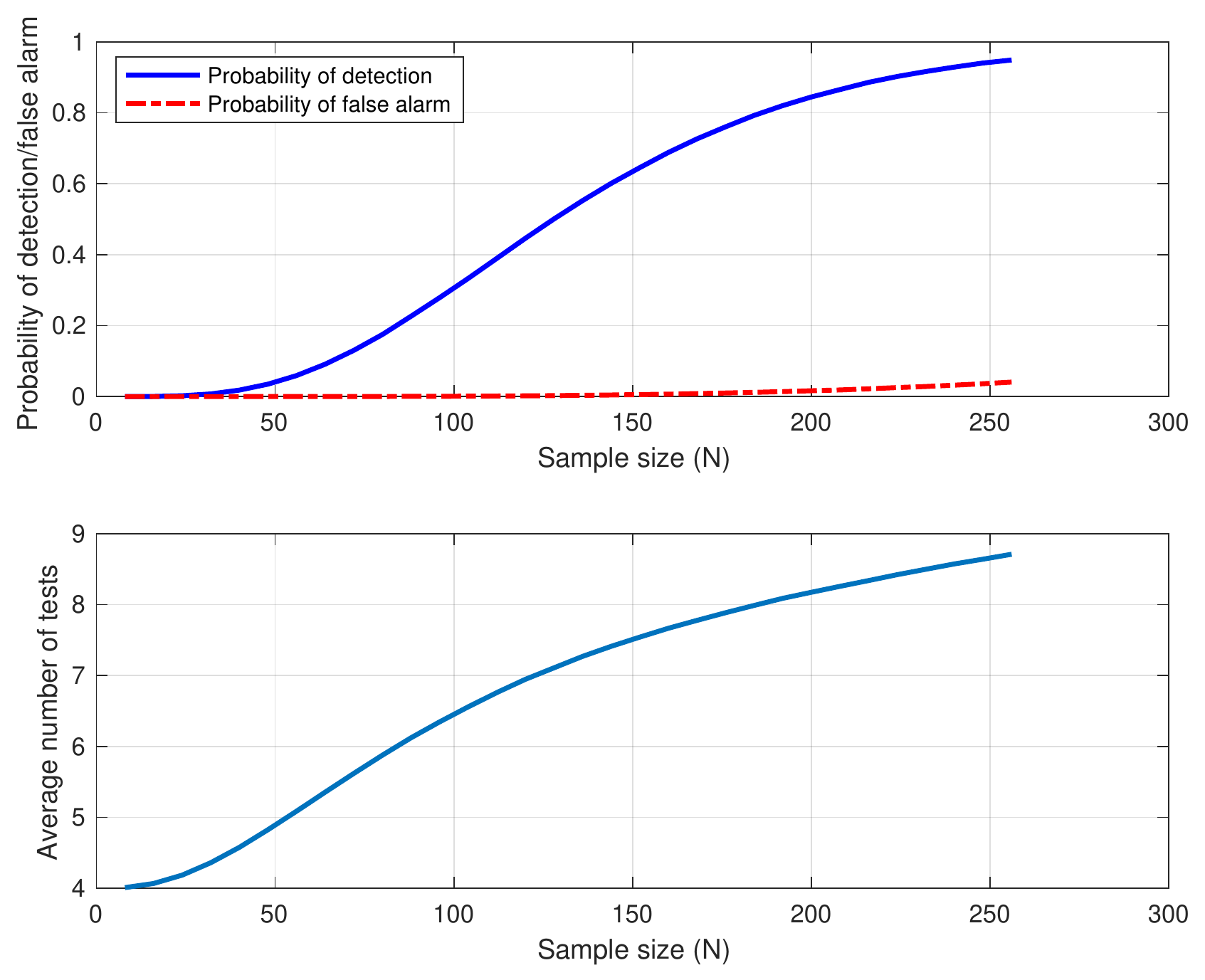}\vspace{-0.25cm}
    \caption{$P_F, P_D$, and $\mathbb{E}[\Gamma]$ as a function of $N$ when $L=8$, $V=4$, $p_0=0.01$, and $p_1 = 0.05$.}
    \label{fig:newresults1}
\end{figure}

Plots of $P_F$ versus $N$, $P_D$ versus $N$, and $\mathbb{E}[\Gamma]$ versus $N$ are shown in Figure~\ref{fig:newresults1}. It can be seen that with less than $8.7$ tests on the average, a probability of detection of $95\%$ can be obtained while the probability of false alarm is only about $4\%$. These results can be improved at the expense of an increase in the average number of tests. 

\begin{figure}[t]
    \centering
    \includegraphics[width=3.75in]{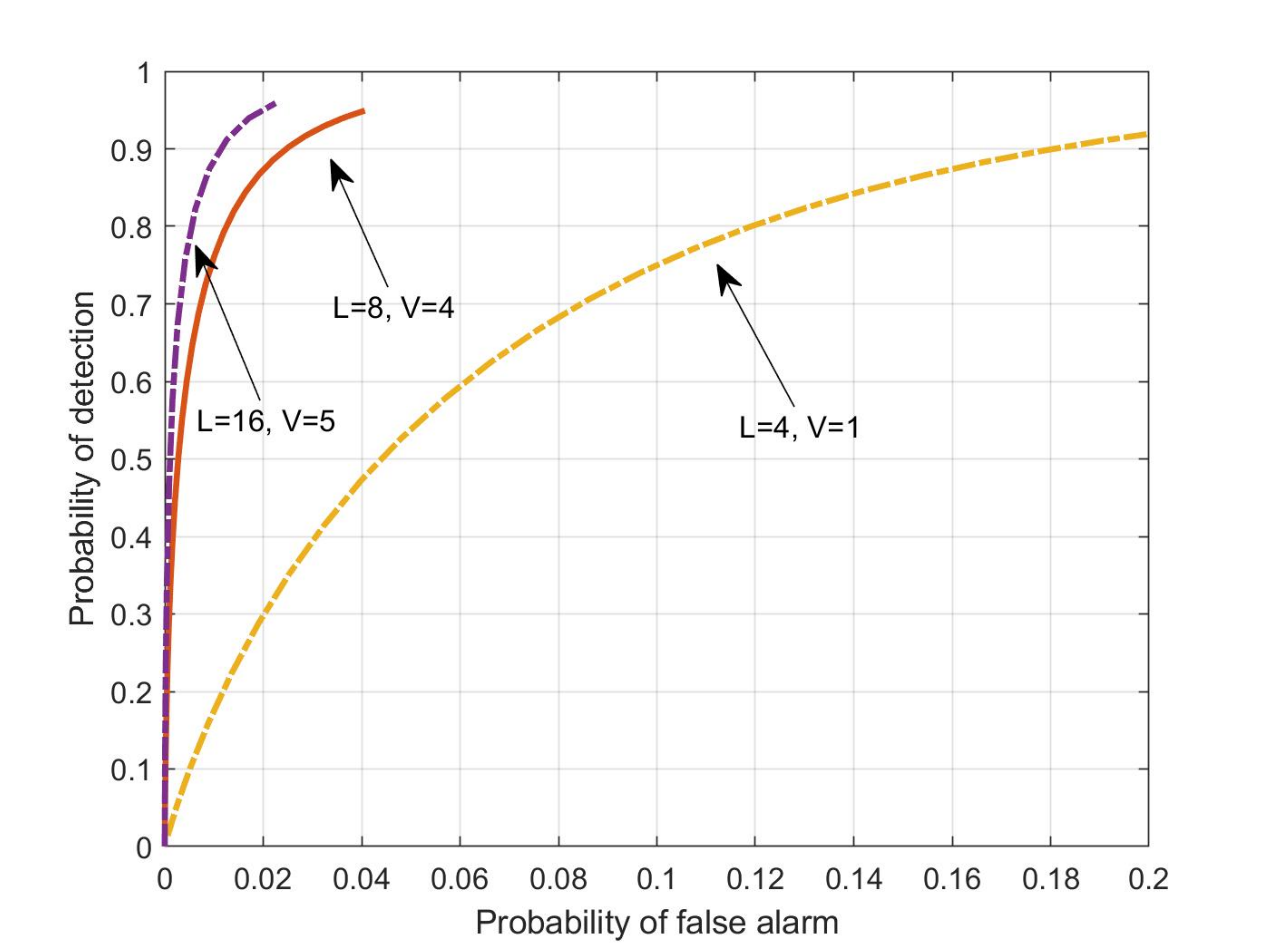}\vspace{-0.25cm}
    \caption{ROC curve obtained by changing $N$ for different values of $L$ and $V$ when $p_0=0.01, p_1 = 0.05$.}
    \label{fig:newresults2}
\end{figure}

In Figure~\ref{fig:newresults2}, we plot the receiver operating characteristics (ROC) curve for different values of $L$ and $V$. (For a given pair of values of $L$ and $V$, each point on the underlying curve corresponds to a different value of $N$.) It can be seen that when $L=16$ and $V=5$, we can obtain an excellent trade-off between $P_F$ and $P_D$. 

\begin{figure}[h]
    \centering
    \includegraphics[width=4.25in]{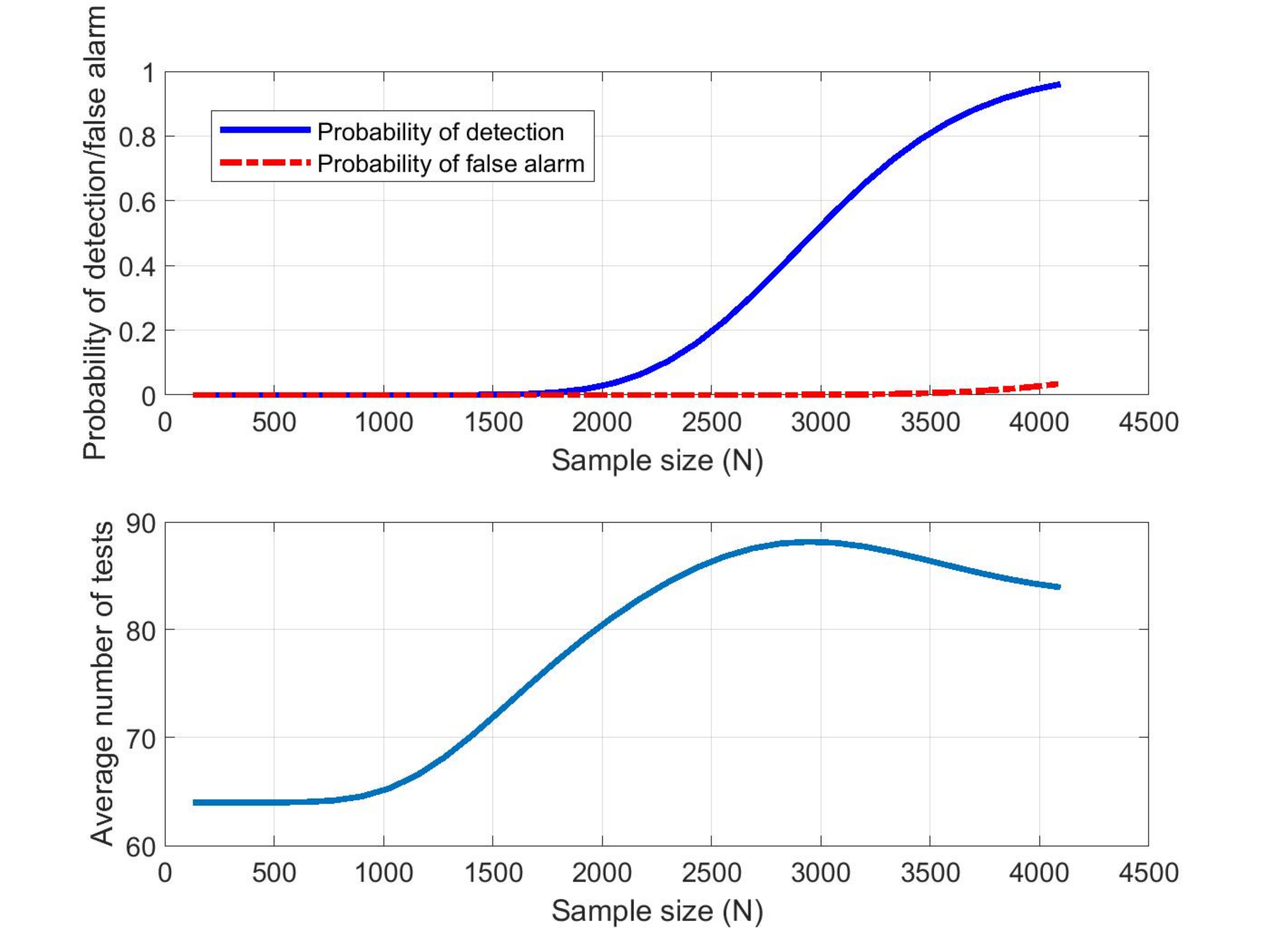}\vspace{-0.25cm}
    \caption{$P_F, P_D$ and $\mathbb{E}[\Gamma]$ as a function of $N$ when $L=128$, $V=26$, $p_0=0.005$, and $p_1 = 0.01$.}
    \label{fig:newresults40961}
\end{figure}

\begin{figure}[h]
    \centering
    \includegraphics[width=3.75in]{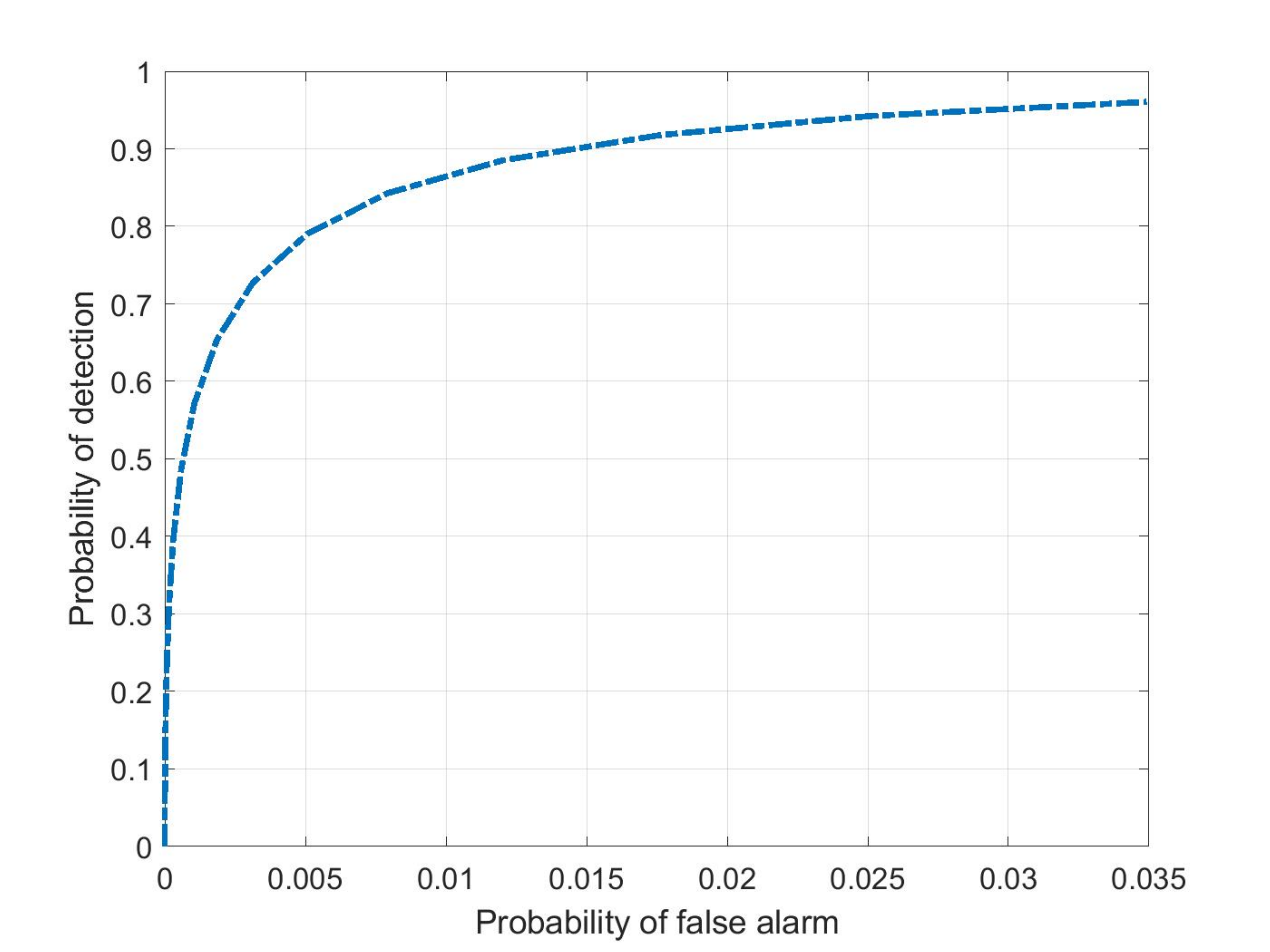}\vspace{-0.25cm}
    \caption{ROC curve obtained by changing $N$ when $L=128$, $V=26$, $p_0=0.005$, and $p_1 = 0.01$.}
    \label{fig:newresults4096ROC}
\end{figure}

When $p_0$ and $p_1$ are closer to each other, more samples are required to distinguish between them. When $p_0 = 0.005$ and $p_1 = 0.01$, we show the $P_F, P_D$ versus $N$ and the average number of tests versus $N$ curves in Figure~\ref{fig:newresults40961} and the ROC curve in Figure~\ref{fig:newresults4096ROC}. In these cases, we have used $L=128$ and $V=26$. It can be seen that when $N=4096$, which corresponds to a maximum pool size of $2N/L=64$, with an average number of tests of $83.9$, a detection probability of $96\%$ and a false alarm probability of $3.5\%$ can be obtained.


\subsubsection{Noisy Setting} 
Due to several factors including human error and current technology, when testing larger subpools the accuracy of tests may be lower than the accuracy of tests on individuals. A recent study in~\cite{Yelin2020} shows that using the standard RT-PCR technology, with a false negative rate of about $10\%$ a single infected individual can be identified in pools of size up to $32$. This naturally raises a question about the robustness of the proposed scheme (for infection rate classification) to the accuracy of tests. In the following, we demonstrate the robustness of the proposed scheme in the presence of noisy test results when the noise is modeled as follows. Given that the hypothesis $H_i$ is true, we assume that the test result for a subpool not containing any infected people will always be negative with probability $1$ (representing a false positive rate of $0\%$, or equivalently, a \emph{test specificity} of $100\%$); whereas the test result for a subpool containing some infected individual(s) will be positive with probability $\rho_i$ (representing a false negative rate of $1-\rho_i$, or equivalently, a \emph{test sensitivity} of $\rho_i$). 

Tables~\ref{table:SmallPools} and~\ref{table:LargePools} show the probability of detection $P_D$, the probability of false alarm $P_F$, and the average number of tests $\mathbb{E}[\Gamma]$ for different values of the pool size $N$, the number of subpools $L$, and the sensitivity $\rho:=\rho_0=\rho_1$. Table~\ref{table:SmallPools} corresponds to the cases with $p_0=0.01$ and $p_1=0.05$, whereas Table~\ref{table:LargePools} corresponds to the cases with $p_0=0.005$ and $p_1=0.01$. For the range of parameters being considered the maximum size of a pool being tested in the proposed scheme (i.e., $2N/L$) is not greater than $64$. 

As can be seen in both tables, when reducing the sensitivity $\rho$ from $100\%$ to $80\%$ (i.e., increasing the level of noise in the test results)  for a wide range of parameters it is still possible to attain $P_D$ and $P_F$ that are within an acceptable range, e.g., $P_D\geq 95\%$ and $P_F\leq 5\%$. For instance, in Table~\ref{table:SmallPools}, for $N=448$ and $L=28$, when $\rho=100\%$ we can achieve $P_D=99.9\%$ and $P_F=4.6\%$, whereas when $\rho=80\%$, $P_D=99.1\%$ and $P_F=4.1\%$ can be achieved. 
It should be noted that in order to achieve (almost) the same $P_D$ and $P_F$ for fixed $N$ and $L$, the threshold $V$ needs to be set to a smaller value as the sensitivity $\rho$ decreases. 
As a result, for a smaller sensitivity $\rho$ the average number of tests required for (almost) the same $P_D$ and $P_F$ is larger. However, the relative increase in the average number of tests becomes smaller for larger values of $N$ and $L$.           

By comparing the results in Tables~\ref{table:SmallPools} and~\ref{table:LargePools}, it can be seen that when the values of $p_0$ and $p_1$ are closer to each other, the pool size $N$ needs to be larger in order to obtain a sufficiently large $P_D$ (about $95\%$) and sufficiently small $P_F$ (about $5\%$). This is expected because for closer values of $p_0$ and $p_1$, distinguishing between the two hypotheses $H_0$ and $H_1$ becomes a more challenging problem, and hence the need for sampling a larger pool of the population. 

\begin{table*}[]
    \centering
    \caption{$P_D$, $P_F$, and $\mathbb{E}[\Gamma]$ for different values of $N$, $L$, and $V$ when $p_0=0.01$ and $p_1 = 0.05$.}
    \begin{tabular}{|c|c|c|c|c|c|c|c|}
    \hline
    Pool size $N$ & No. of subpools $L$ & Max. pool size $\frac{2N}{L}$ & Sensitivity $\rho$ & $P_D$ & $P_F$ & $\mathbb{E}[\Gamma]$ & Threshold $V$ \\
    \hline
    256 & 8 & 64 & 100\% & 96.0\% & 4.1\% & 8.7 & 4 \\
    \hline
    256 & 8 & 64 & 80\% & 88.8\% & 7.7\% & 6.7 & 3 \\
    \hline
    256 & 16 & 32 & 100\% & 98.8\% & 7.6\% & 9.3 & 4 \\
    \hline
    256 & 16 & 32 & 80\% & 91.2\% & 3.4\% & 9.8 & 4 \\
    \hline
    448 & 14 & 64 & 100\% & 99.8\% & 6.2\% & 10.4 & 6 \\
    \hline
    448 & 14 & 64 & 80\% & 97.3\% & 6.6\% & 11.1 & 5 \\
    \hline
    448 & 28 & 32 & 100\% & 99.9\% & 4.6\% & 16.8 & 7 \\
    \hline
    448 & 28 & 32 & 80\% & 99.1\% & 4.1\%  & 16.1 & 6 \\
    \hline 
    \end{tabular}
    \label{table:SmallPools}
\end{table*}

\begin{table*}[]
    \centering
    \caption{$P_D$, $P_F$, and $\mathbb{E}[\Gamma]$ for different values of $N$, $L$, and $V$ when $p_0=0.005$ and $p_1 = 0.01$.}
    \begin{tabular}{|c|c|c|c|c|c|c|c|}
    \hline
    Pool size $N$ & No. of subpools $L$ & Max. pool size $\frac{2N}{L}$ & Sensitivity $\rho$ & $P_D$ & $P_F$ & $\mathbb{E}[\Gamma]$ & Threshold $V$ \\
    \hline
     4096 & 128 & 64 & 100\% & 97.5\% & 5.6\% & 83.0 & 25 \\
     \hline
     4096 & 128 & 64 & 80\% & 92.5\% & 4.7\% & 81.0 & 21 \\
     \hline
     4096 & 256 & 32 & 100\% & 98.3\% & 6.1\% & 146.0 & 26 \\
     \hline
     4096 & 256 & 32 & 80\% & 94.2\% & 4.6\% & 143.2 & 22 \\
     \hline
     4736 & 148 & 64 & 100\% & 98.3\% & 4.4\% & 95.8 & 29 \\
     \hline
     4736 & 148 & 64 & 80\% & 94.9\% & 4.3\% & 92.7 & 24 \\
     \hline
     4736 & 296 & 32 & 100\% & 98.9\% & 5.2\% & 168.9 & 30 \\
     \hline
     4736 & 296 & 32 & 80\% & 96.4\% & 4.6\% & 165.6 & 25 \\
    \hline 
    \end{tabular}
    \label{table:LargePools}
\end{table*}

\bibliographystyle{IEEEtran}
\bibliography{journal_abbr,sparseestimation,grouptesting,MACcollision,0-Bibliography,COVID-19}

\end{document}